\documentclass[epj]{webofc}
\usepackage[utf8]{inputenc}
\usepackage[varg]{txfonts}   
\usepackage{booktabs}
\usepackage{xcolor}
\usepackage{amsmath}
\usepackage{amssymb}
\usepackage{amsfonts}
\usepackage{bm}
\usepackage{bbold}
\usepackage{slashed}
\definecolor{darkred}{rgb}{0.4,0.0,0.0}
\definecolor{darkgreen}{rgb}{0.0,0.4,0.0}
\definecolor{darkblue}{rgb}{0.0,0.0,0.4}
\usepackage[bookmarks,linktocpage,colorlinks,
    linkcolor = darkred,
    urlcolor  = darkblue,
    citecolor = darkgreen]{hyperref}
\usepackage{subfigure}
\wocname{EPJ Web of Conferences}
\woctitle{Lattice2017}
\begin{document}
%
\selectlanguage{english}
\title{%
Perturbative matching of \\ continuum and lattice quasi-distributions
}
\author{%
\firstname{Tomomi} \lastname{Ishikawa}\inst{1}\fnsep\thanks{Speaker, \email{tomomi.ik@gmail.com}}
}
\institute{%
T.~D.~Lee Institute, Shanghai Jiao Tong University
Shanghai, 200240, P. R. China
}
\abstract{%
Matching of the quasi parton distribution functions between continuum and lattice is addressed using lattice perturbation theory specifically with Wilson-type fermions. The matching is done for nonlocal quark bilinear operators with a straight Wilson line in a spatial direction. We also investigate operator mixing in the renormalization and possible ${\cal O}(a)$ operators for the nonlocal operators based on a symmetry argument on lattice.
}
\maketitle
\section{Introduction}\label{intro}

Understanding the internal structure of nucleons through quantum chromodynamics (QCD) gives phenomenological implications to high-energy physics and astrophysics. Especially, determination of parton distribution functions (PDFs) is significantly essential to bridge energy scales between the fundamental degree of freedom, quarks and gluons, and the hadrons, such as pions and protons. In studying high energy scattering processes, one of the key concepts is the ``QCD collinear factorization'' to separate the perturbative part from the nonperturbative physics. The scattering cross sections are written in a convolution of the perturbative hard part and nonperturbative PDFs, which absorb all collinear divergences of the partonic scattering. Using the factorization, the determination of the PDFs are mostly carried out through global QCD analyses, where experimental data are combined with perturbative hard parts.

Direct lattice QCD calculation of the PDFs is desirable to obtain complementary information to the global QCD analysis. However, time-dependent quantities cannot be directly treated on the lattice, and for this reason, it is difficult to calculate the PDFs defined in the light-cone coordinate. Traditional treatment to calculate the PDFs on the lattice is to use its Mellin moments, which is time-independent. By the Mellin transformation, PDFs can be in principle reconstructed from the moments. However, the moments are hard to be accessed except for first few moments due to power-divergent mixing between operators. Recently, quasi-PDFs approach was proposed to overcome the difficulty in calculating the PDFs on the lattice~\cite{Ji:2013dva}. The quasi-(quark) PDFs are defined by
\begin{eqnarray}
\widetilde{q}(\tilde{x},\tilde{\mu}, P_z)=
\int\frac{d\delta z}{4\pi}e^{-i\delta z\tilde{x}P_z}
\langle P_z|\overline{\psi}(\delta z)\gamma_3
\exp\left(-ig\int_0^{\delta z}dz'A_3(z')\right)\psi(0)| P_z\rangle,
\label{EQ:quasi-quark-PDFs}
\end{eqnarray}
which is Fourier transform of in a coordinate space nucleon matrix elements with momentum in $z$-direction, $P_z$. Because the two quark fields in the nonlocal quark bilinear are separated in a purely spatial direction ($z$-direction), the quasi-PDFs are calculable on the lattice. The quasi-PDFs are related with the light-cone PDFs $q(x,\mu)$ by the large momentum effective theory~\cite{Ji:2014gla}:
\begin{eqnarray}
\widetilde{q}(x,\tilde{\mu}, P_z)&=&
Z\left(x,\frac{\tilde{\mu}}{P_z}, \frac{\mu}{P_z}\right)\otimes q(x,\mu)
+{\cal O}\left(\frac{\Lambda_{\rm QCD}^2}{P_z^2}, \frac{M^2}{P_z^2}\right),
\label{EQ:LaMET_matching}
\end{eqnarray}
where $\otimes$ represents a convolution with respect to $x$, and $M$ is a nucleon mass. The $Z$ factor in Eq.~(\ref{EQ:LaMET_matching}) can be perturbatively obtained. The similar approaches have been proposed in Refs.~\cite{Ma:2014jla, Ma:2017pxb, Radyushkin:2017cyf, Orginos:2017kos}.

The renormalization of the nonlocal quark bilinear in Eq.~(\ref{EQ:quasi-quark-PDFs}),
\begin{eqnarray}
O_{\Gamma}(\delta z)=
\overline{\psi}(x+\hat{\bm{3}}\delta z)\Gamma
U_{3}(x+\hat{\bm{3}}\delta z; x)\psi(x),
\label{EQ:nonlocal_bilinear}
\end{eqnarray}
has been perturbatively studied since 1980s~\cite{Arefeva1980, Dotsenko:1979wb, Craigie:1980qs, Dorn:1986dt, Stefanis:1983ke}. Since the nonlocal operator involves a Wilson line $U_{3}(x+\hat{\bm{3}}\delta z; x)$, it intrinsically suffers from ultraviolet (UV) power divergences. Taking the power divergence into account, the renormalization pattern is written as
\begin{eqnarray}
O_{\Gamma}^{{\rm ren}(\mu)}(\delta z)=e^{\delta m(\mu)|\delta z|}Z_{\psi, z}(\delta z, \mu)O_{\Gamma}(\delta z),
\end{eqnarray} 
where $\delta m$ is the mass renormalization of a test particle moving along the Wilson line, which contains the power divergence. $Z_{\psi, z}(\delta z, \mu)$ removes the remaining logarithmic divergences. The all-order proof of the renormalizability in this pattern has been reported in Refs.~\cite{Ji:2017oey, Ishikawa:2017faj}. The nonperturbative subtraction method of the power divergence in the nonlocal bilinear has been proposed in Refs.~\cite{Musch:2010ka, Ishikawa:2016znu, Chen:2016fxx}. Recently, the nonperturbative renormalization of this operator using RI/MOM scheme has been demonstrated in Refs.~\cite{Alexandrou:2017huk, Chen:2017mzz, Green:2017xeu, Stewart:2017tvs}.

Lattice QCD formulation respects not all the symmetries possessed in the continuum theory. Lacking some symmetries makes the renormalization pattern different from that in the continuum. Notably, the nonlocal quark bilinear (\ref{EQ:nonlocal_bilinear}) can mix with different $\Gamma$ on the lattice, which was found in the one-loop lattice perturbative calculation~\cite{Constantinou:2017sej} and later also supported by Ref.~\cite{Green:2017xeu}. In this proceedings, we address the operator mixing from both the lattice perturbation and a view of lattice action symmetry which is a nonperturbative argument. The discussion using the symmetry is also extended to check ${\cal O}(a)$ operators for the nonlocal quark bilinears~\cite{Chen:2017mie}.

\section{One-loop perturbative matching between continuum and lattice with Wilson fermion}\label{SEC1}

\begin{figure}[tb]
\centering
\parbox{34mm}{
\centering
\includegraphics[scale=0.26, viewport = 0 0 320 300, clip]
{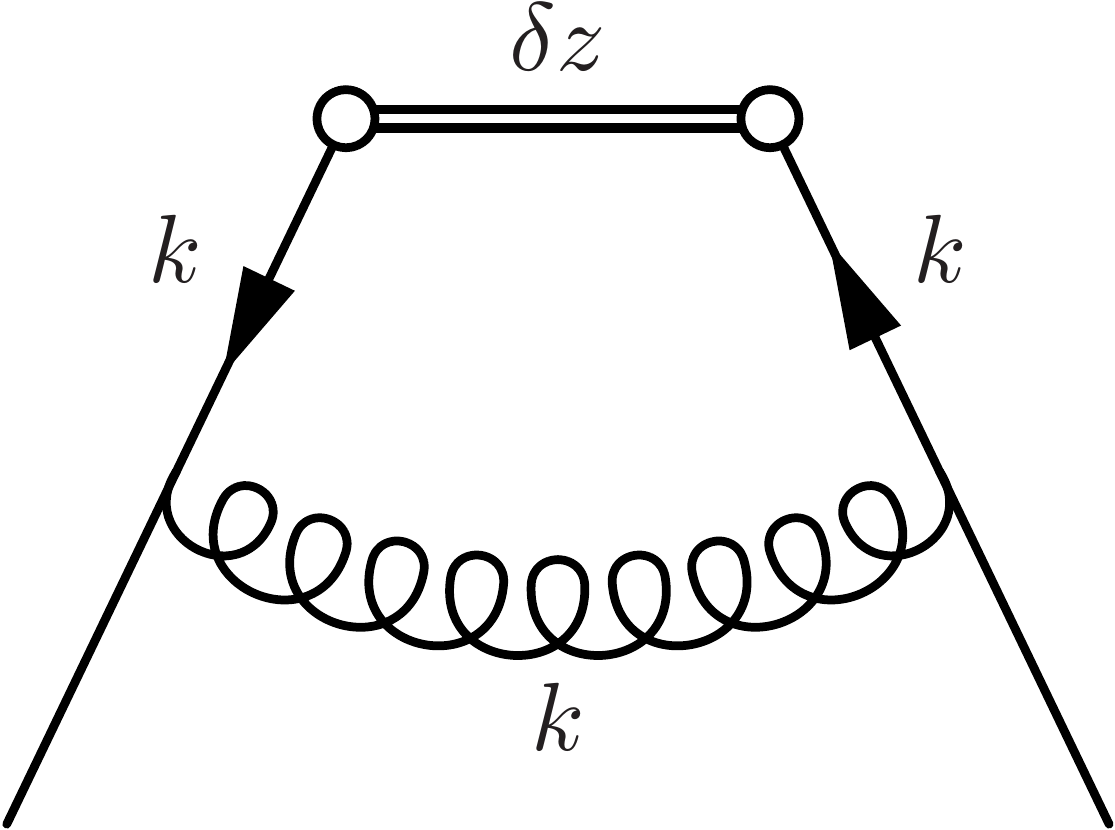}
$\delta\Gamma_{\rm vertex}(\delta z)$
}
\parbox{68mm}{
\centering
\includegraphics[scale=0.26, viewport = 0 0 320 300, clip]
{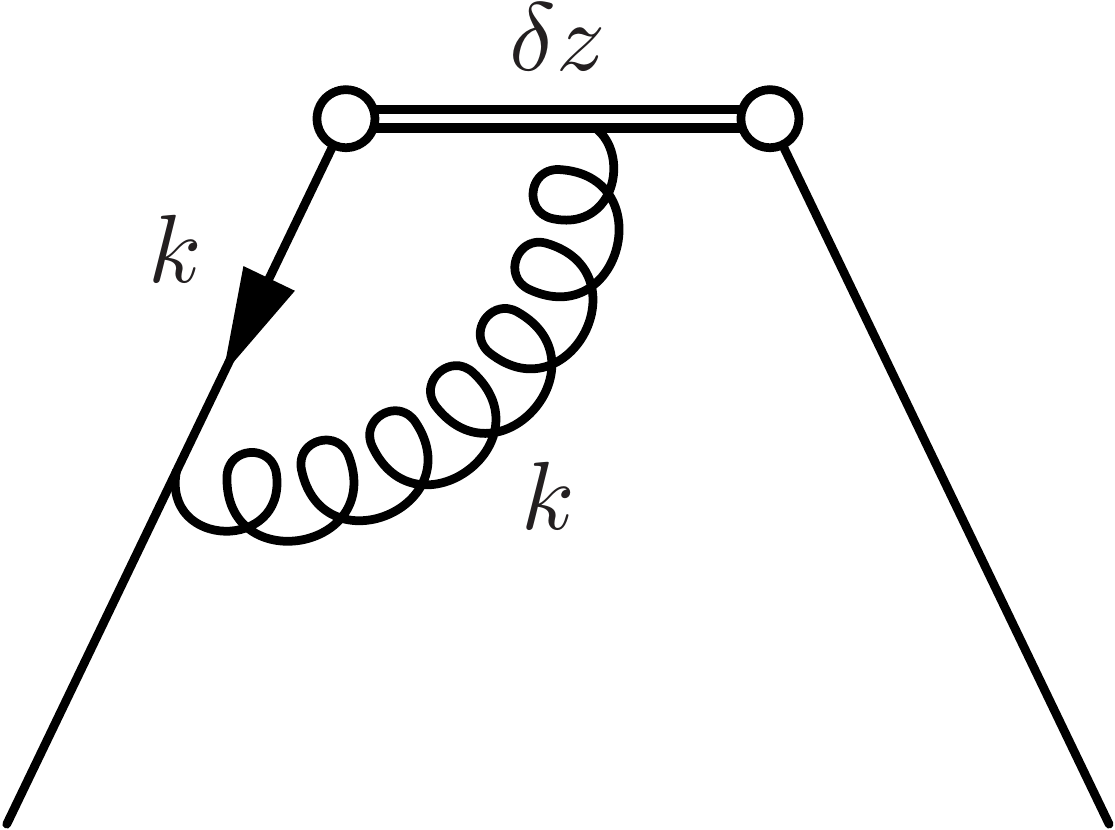}
\includegraphics[scale=0.26, viewport = 0 0 320 300, clip]
{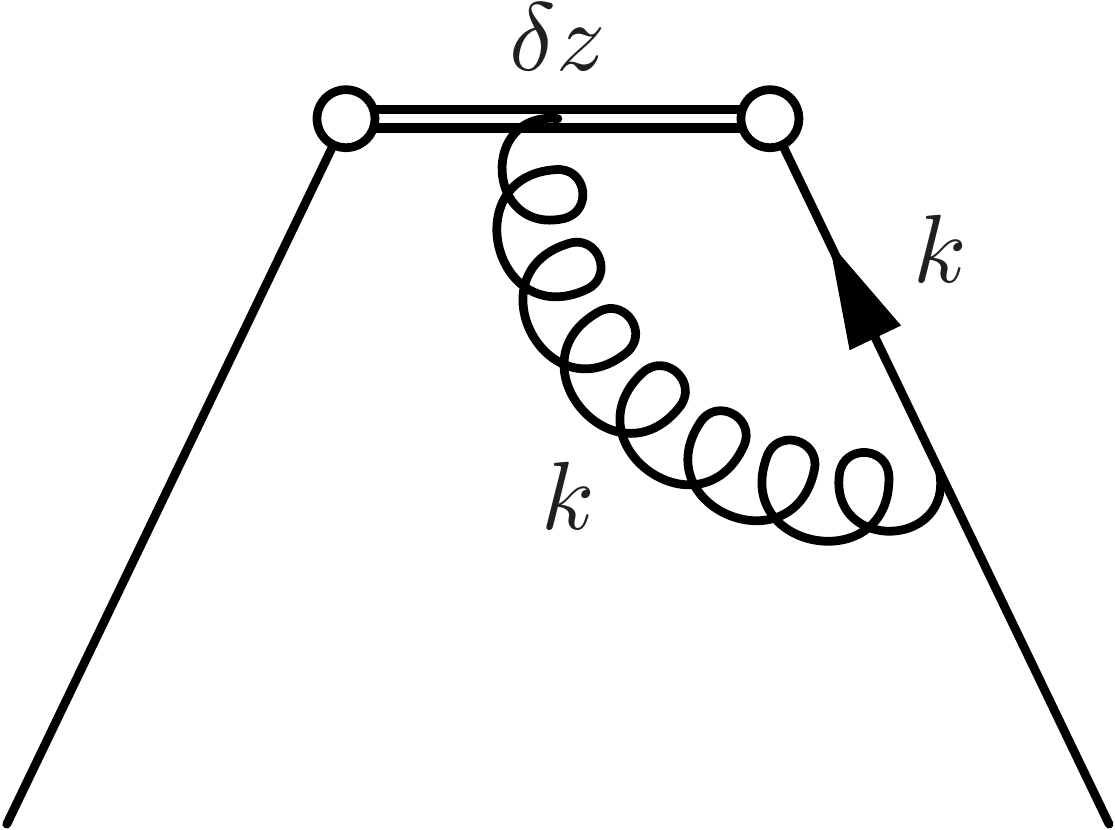} 
$\delta\Gamma_{\rm sail}(\delta z)$
}
\parbox{34mm}{
\centering
\includegraphics[scale=0.26, viewport = 0 0 320 320, clip]
{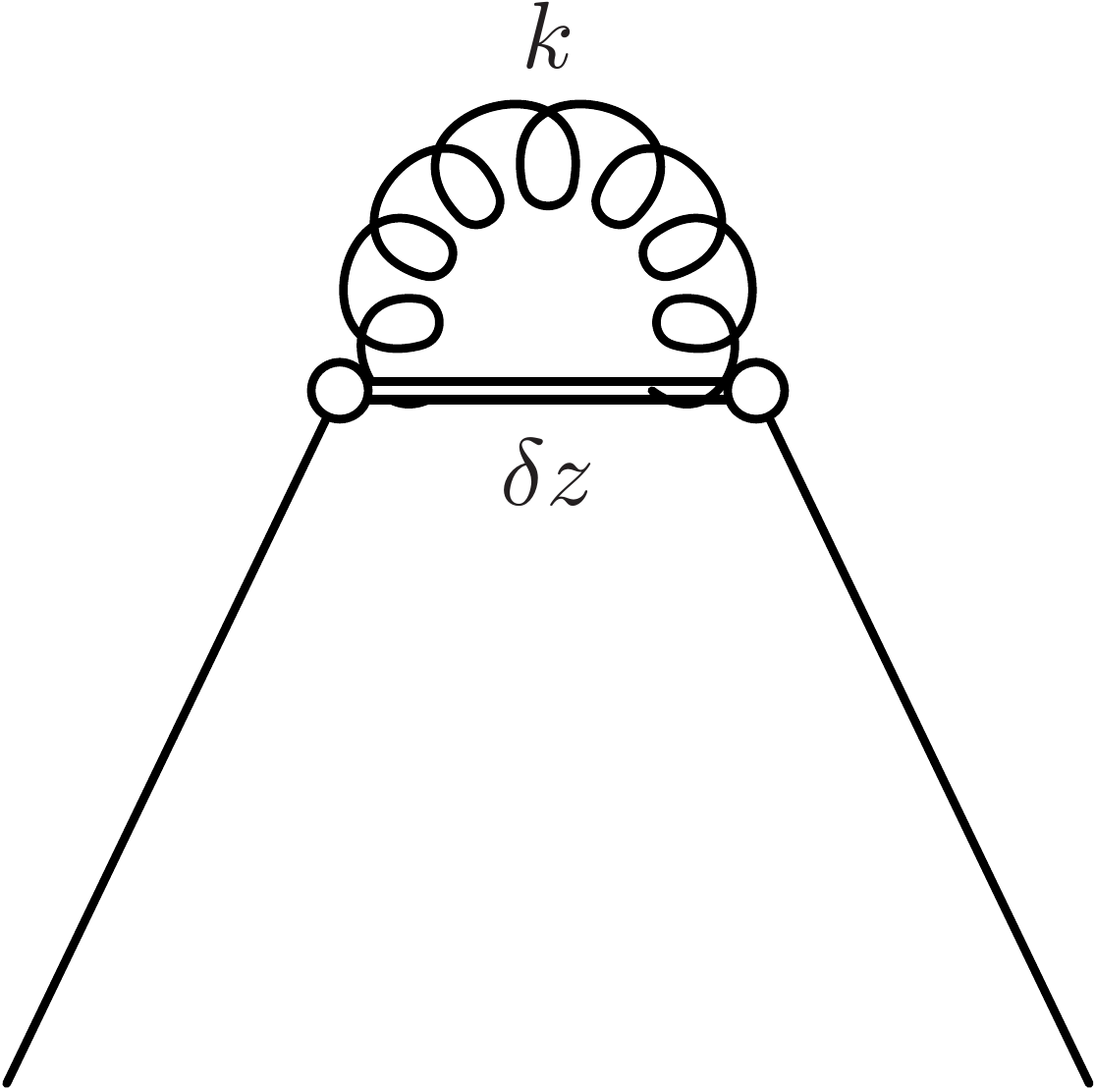} 
$\delta\Gamma_{\rm tadpole}(\delta z)$
 }
\caption{One-loop diagrams for the nonlocal quark bilinear.}
\label{FIG:diagrams} 
\end{figure}

\begin{figure}[tb]
\centering
\parbox{70mm}{
\centering
unsmear\\
\parbox{34mm}{
\centering
$c_{\gamma_3\gamma_3}(\delta z)$
\includegraphics[scale=0.48, viewport = 0 0 200 470, clip]
{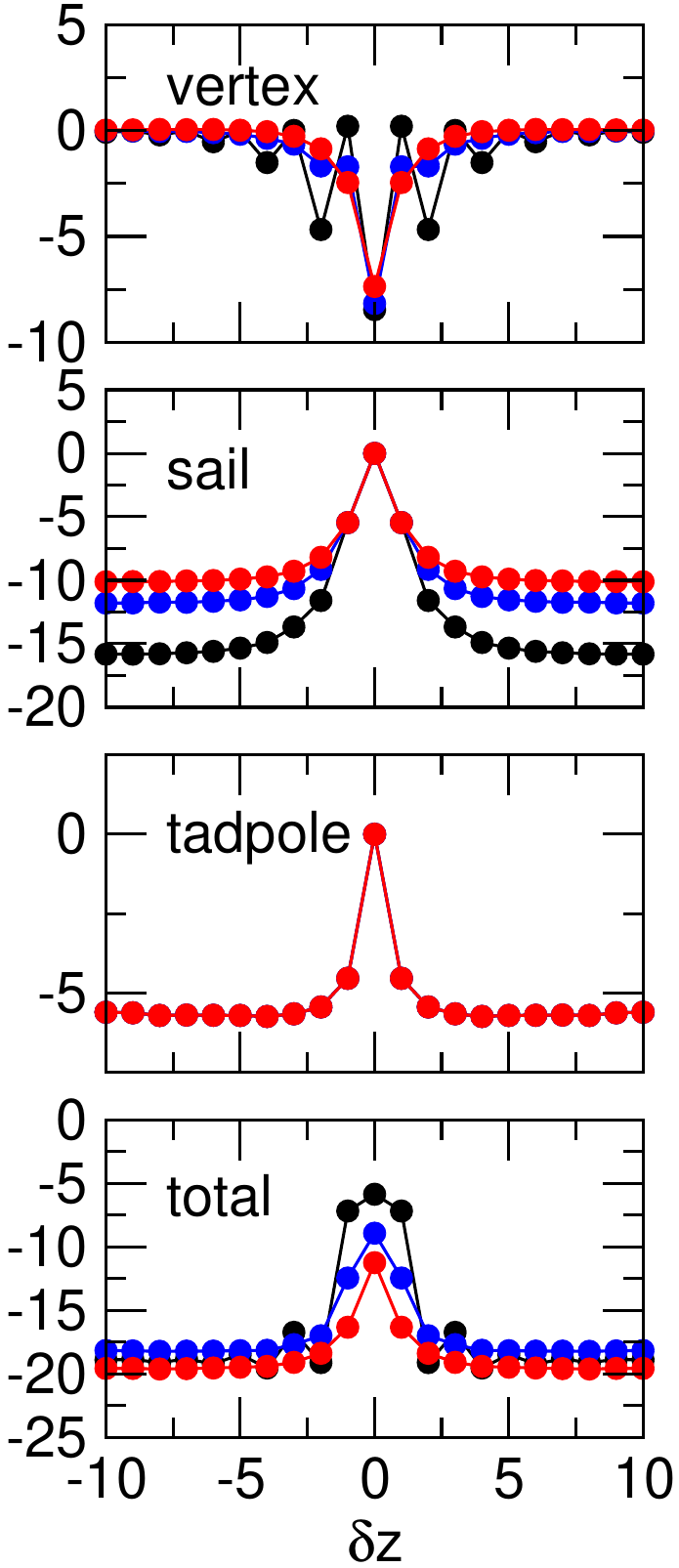}
}
\parbox{34mm}{
\centering
$c_{\gamma_3\mathbb{1}}(\delta z)$
\includegraphics[scale=0.48, viewport = 0 0 200 470, clip]
{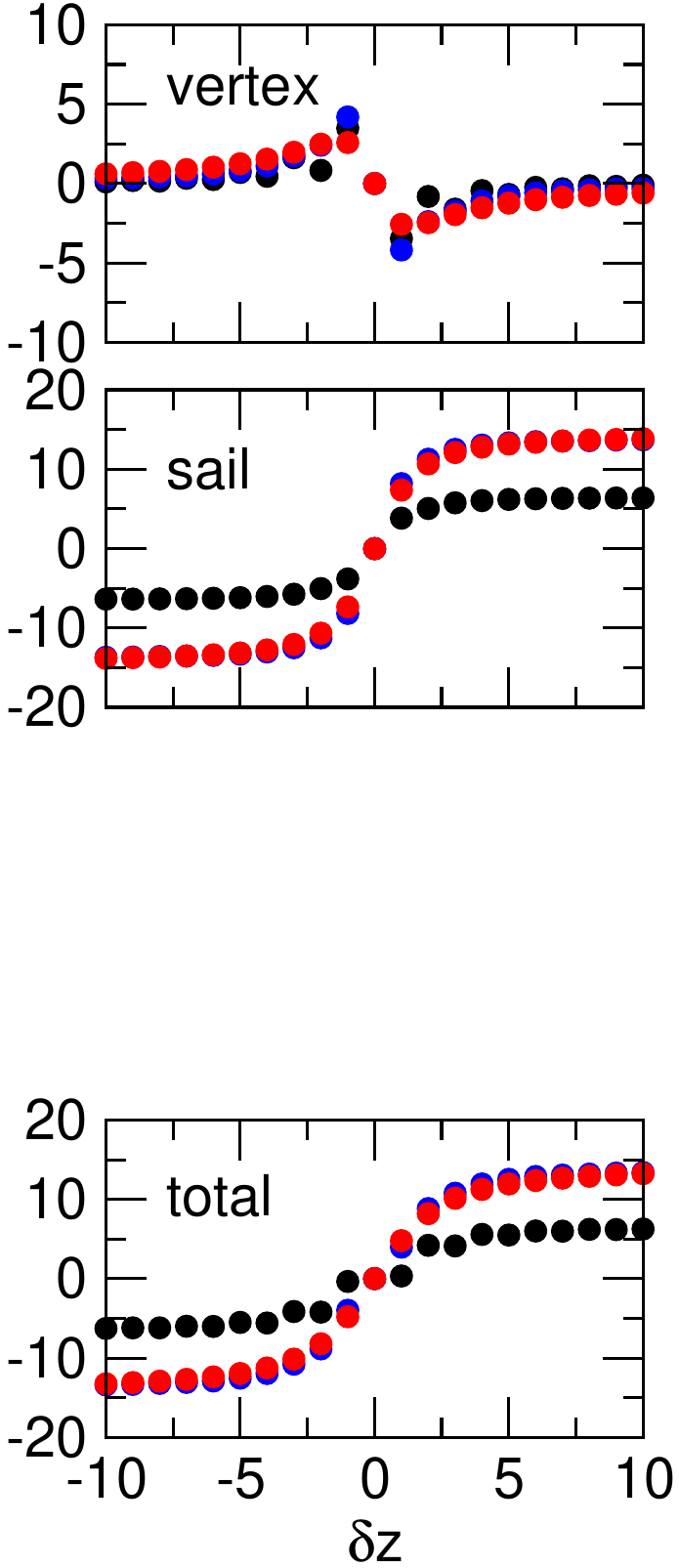}
}
}
\parbox{70mm}{
\centering
HYP2\\
\parbox{34mm}{
\centering
$c_{\gamma_3\gamma_3}(\delta z)$
\includegraphics[scale=0.48, viewport = 0 0 200 470, clip]
{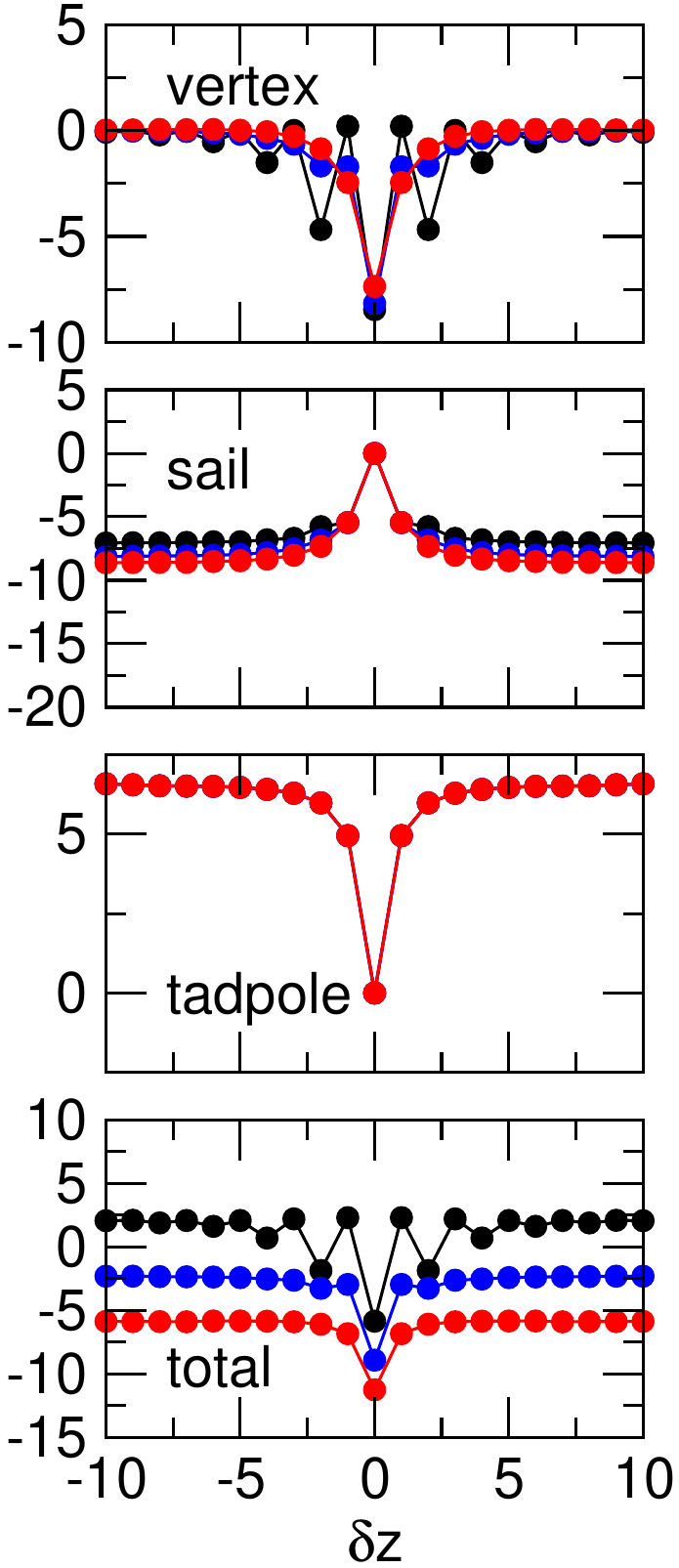} 
}
\parbox{34mm}{
\centering
$c_{\gamma_3\mathbb{1}}(\delta z)$
\includegraphics[scale=0.48, viewport = 0 0 200 470, clip]
{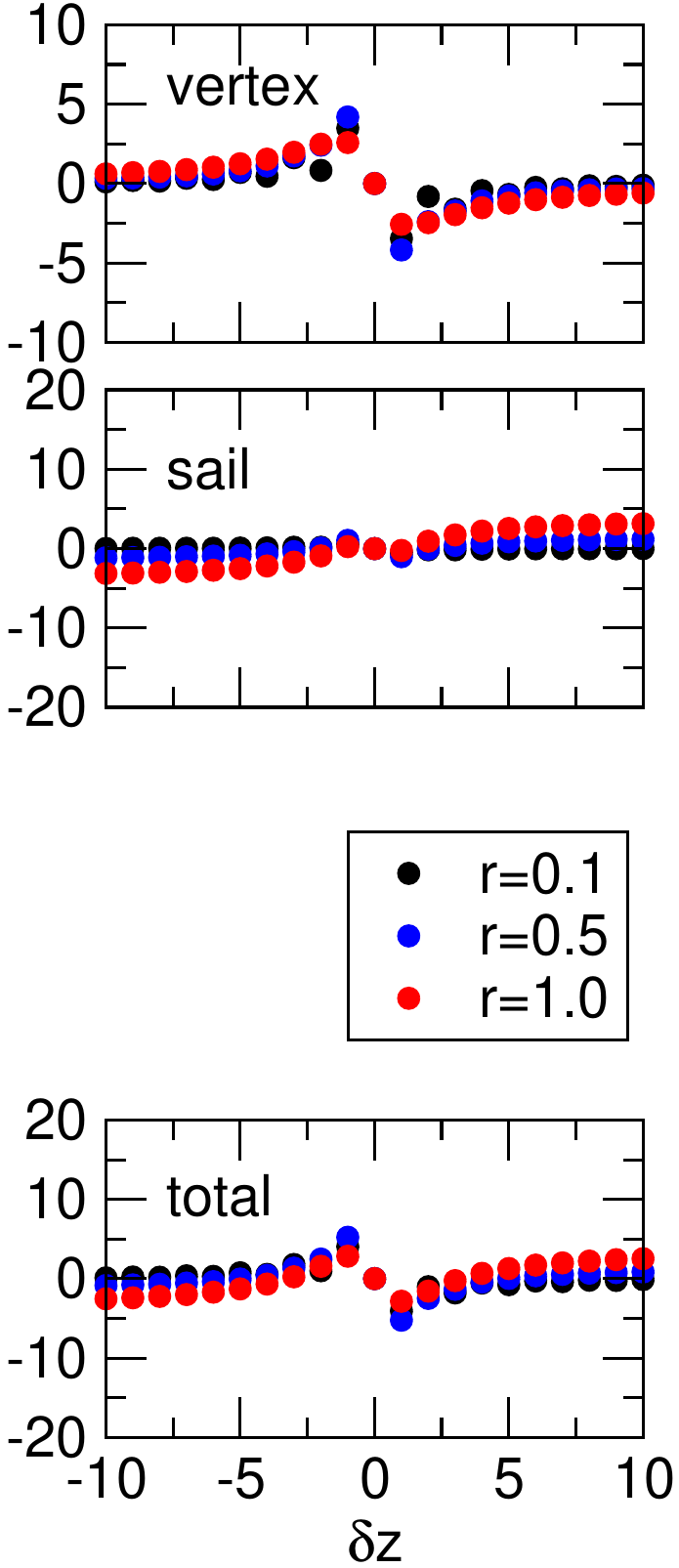}
}
}
\caption{One-loop matching coefficients for each diagram: vertex-type, sail-type, and tadpole-type, as well as their total contribution (including wave function part). The linear divergence is subtracted, and the MF improvement~\cite{Ishikawa:2016znu} is used. Left figures show the results for unsmeared Wilson line, while right ones show for HYP2 smeared case.}
\label{FIG:one-loop_matching_ coeff} 
 \end{figure}

In this section, we investigate the matching factor for the nonlocal quark bilinear (\ref{EQ:nonlocal_bilinear}) between continuum and lattice using one-loop perturbation theory. In the calculation, we assume the Wilson fermion formalism with a plaquette gluon action. The value of Wilson parameter $r$ is kept unspecified, and Feynman gauge is used for the gauge fixing. In the continuum side, we set three dimensional UV cutoff in the direction perpendicular to $z$ to regulate the UV divergence, for simplicity. The detail of this scheme can be seen in Ref.~\cite{Ishikawa:2016znu}. The one-loop diagrams we calculate here are shown in Fig.~\ref{FIG:diagrams}; from left, vertex-type, sail-type, and tadpole-type diagram. By the lattice perturbation theory, the one-loop amplitude is written in a form:
\begin{eqnarray}
\begin{bmatrix}
\langle O_{\Gamma}(\delta z)\rangle \\ \langle O_{\Gamma'}(\delta z)\rangle
\end{bmatrix}
=
\mathbb{1}+g^2
\begin{bmatrix}
{\cal A}_{\Gamma\Gamma}(\delta z, r^2) &
r{\cal A}_{\Gamma\Gamma'}(\delta z, r^2) \\
r{\cal A}_{\Gamma'\Gamma}(\delta z, r^2) &
{\cal A}_{\Gamma'\Gamma'}(\delta z, r^2) \\
\end{bmatrix}
\begin{bmatrix}
\langle O_{\Gamma}(\delta z)\rangle_{\rm tree} \\
\langle O_{\Gamma'}(\delta z)\rangle_{\rm tree}
\end{bmatrix}
+{\cal O}(g^4),
\end{eqnarray}
where ${\cal A}_{\Gamma\Gamma}$, ${\cal A}_{\Gamma\Gamma'}$, ${\cal A}_{\Gamma'\Gamma}$, and ${\cal A}_{\Gamma'\Gamma'}$ are one-loop coefficients. The operator $O_{\Gamma}(\delta z)$ can mix with $O_{\Gamma'=\gamma_3\Gamma+\Gamma\gamma_3}(\delta z)$, which is the same structure as that originally obtained in Ref.~\cite{Constantinou:2017sej} for the clover-Wilson fermion. The mixing part is always proportional to $r$, and thus $r=0$ case does not cause the operator mixing~\cite{Ishikawa:2016znu}, which indicates the mixing is due to lack of chiral symmetry in the Wilson fermion formalism.
In the calculation, we also consider the case where the Wilson line is smeared with HYP2 smearing~\cite{DellaMorte:2005nwx}. The one-loop coefficients $c(\delta z, r)$ in the matching between continuum and lattice,
\begin{eqnarray}
\begin{bmatrix}
\langle O_{\Gamma}(\delta z)\rangle^{\rm cont} \\
\langle O_{\Gamma'}(\delta z)\rangle^{\rm cont}
\end{bmatrix}
=
\mathbb{1}+\frac{g^2}{(4\pi)^2}C_F
\begin{bmatrix}
c_{\Gamma\Gamma}(\delta z, r)  & c_{\Gamma\Gamma'}(\delta z, r) \\
c_{\Gamma'\Gamma}(\delta z, r) & c_{\Gamma'\Gamma'}(\delta z, r) \\
\end{bmatrix}
\begin{bmatrix}
\langle O_{\Gamma}(\delta z)\rangle^{\rm latt} \\
\langle O_{\Gamma'}(\delta z)\rangle^{\rm latt}
\end{bmatrix}
+{\cal O}(g^4), 
\end{eqnarray}
are calculated for $\Gamma=\gamma_3$ and $\Gamma'=\mathbb{1}$, and shown in Fig.~\ref{FIG:one-loop_matching_ coeff}. In this case, there is a mixing. As we can see, the smearing on the Wilson line makes the operator mixing negligible in the larger $\delta z$ region.

\section{Symmetry analysis for the operator mixing}\label{SEC2}

The operator mixing pattern seen in the one-loop perturbative calculation can be confirmed at nonperturbative level using action symmetries. The symmetries used here are chiral symmetry ($\chi$) and discrete symmetries: parity ($\cal P$), time reversal ($\cal T$), and charge conjugation ($\cal C$). For some lattice fermion actions, such as Wilson fermion, the chiral symmetry is explicitly broken.

We here summarize the symmetry transformation in Euclidean spaces~\cite{Gattringer:2010zz} used in the discussion, where three spatial and Euclidean-time directions are denoted by $(x, y, z, t)$ or $(1, 2, 3, 4)$.
\begin{itemize}
\item Parity ($\cal P$):

In Euclidean spaces, a parity transformation can be defined for any direction. The general parity transformation ${\cal P}_{\mu}$ is
\begin{eqnarray}
\psi(x)&\xrightarrow[]{{\cal P}_{\mu}}&
\psi(x)^{{\cal P}_{\mu}}=\gamma_{\mu}\psi(\mathbb{P}_{\mu}(x)),\\
\overline{\psi}(x)&\xrightarrow[]{{\cal P}_{\mu}}&
\overline{\psi}(x)^{{\cal P}_{\mu}}=\overline{\psi}(\mathbb{P}_{\mu}(x))\gamma_{\mu},\\
U_{\nu\not=\mu}(x)&\xrightarrow[]{{\cal P}_{\mu}}&
U_{\nu\not=\mu}(x)^{{\cal P}_{\mu}}
=U_{\nu\not=\mu}^{\dagger}(\mathbb{P}_{\mu}(x)-\hat{\nu}),\\
U_{\mu}(x)&\xrightarrow[]{{\cal P}_{\mu}}&
U_{\mu}(x)^{{\cal P}_{\mu}}=U_{\mu}(\mathbb{P}_{\mu}(x)),
\end{eqnarray}
where $\mathbb{P}_{\mu}(x)$ is the vector $x$ with sign flipped except for the $\mu$-direction.
\item Time reversal ($\cal T$):

In Euclidean spaces, a time reversal can be generalized in any direction. The general time reversal ${\cal T}_{\mu}$ is
\begin{eqnarray}
\psi(x)&\xrightarrow[]{{\cal T}_{\mu}}&
\psi(x)^{{\cal T}_{\mu}}=\gamma_{\mu}\gamma_5\psi(\mathbb{T}_{\mu}(x)),\\
\overline{\psi}(x)&\xrightarrow[]{{\cal T}_{\mu}}&
\overline{\psi}(x)^{{\cal T}_{\mu}}=
\overline{\psi}(\mathbb{T}_{\mu}(x))\gamma_5\gamma_{\mu},\\
U_{\mu}(x)&\xrightarrow[]{{\cal T}_{\mu}}&
U_{\mu}(x)^{{\cal T}_{\mu}}=U_{\mu}^{\dagger}(\mathbb{T}_{\mu}(x)-\hat{\mu}),\\
U_{\nu\not=\mu}(x)&\xrightarrow[]{{\cal T}_{\mu}}&
U_{\nu\not=\mu}(x)^{{\cal T}_{\mu}}
=U_{\nu\not=\mu}(\mathbb{T}_{\mu}(x)),
\end{eqnarray}
where $\mathbb{T}_{\mu}(x)$ is the vector $x$ with sign flipped in the $\mu$-direction.
\item Charge conjugation ($\cal C$):

Charge conjugation ${\cal C}$ transforms particles into antiparticles and it is
expressed as
\begin{eqnarray}
\psi(x)&\xrightarrow[]{\cal C}&
\psi(x)^{\cal C}=C^{-1}\overline{\psi}(x)^{\top},\\
\overline{\psi}(x)&\xrightarrow[]{\cal C}&
\overline{\psi}(x)^{\cal C}=-\psi(x)^{\top}C,\\
U_{\mu}(x)&\xrightarrow[]{\cal C}&
U_{\mu}(x)^{\cal C}=U_{\mu}(x)^{\ast}=(U_{\mu}^{\dagger}(x))^{\top},
\end{eqnarray}
where the charge conjugation matrix $C$ obeys the relation
\begin{eqnarray}
C\gamma_{\mu}C^{-1}=-\gamma_{\mu}^{\top}, \;\;\;\;\;
C\gamma_5C^{-1}=\gamma_5^{\top}.
\end{eqnarray}
\item Chiral rotation ($\chi$):

Chiral rotation of the fermion fields $\chi$ is presented as
\begin{eqnarray}
\psi(x)&\xrightarrow[]{\chi}&\psi'(x)=e^{i\alpha\gamma_5}\psi(x),\\
\overline{\psi}(x)&\xrightarrow[]{\chi}&\overline{\psi}'(x)
=\overline{\psi}(x)e^{i\alpha\gamma_5},
\label{EQ:chiral_transformation}
\end{eqnarray}
where $\alpha$ represents a rotation parameter. With the existence of quark mass $m$, chiral symmetry in the action is softly broken. The effect of the nonzero quark mass is analyzed by introducing a spurious chiral transformation
\begin{eqnarray}
m\xrightarrow[]{\chi'}e^{-i\alpha\gamma_5}me^{-i\alpha\gamma_5},
\end{eqnarray}
so that the quark mass term is invariant under the transformation.
\end{itemize}

Because the nonlocal quark bilinear
\begin{eqnarray}
O_{\Gamma}(\delta z)=
\overline{\psi}(x+\hat{\bm{3}}\delta z)\Gamma
U_{3}(x+\hat{\bm{3}}\delta z; x)\psi(x),
\label{EQ:nonlocal_bilinear2}
\end{eqnarray}
has a specific direction, $z$, we take this into account in treating Dirac matrices $\Gamma$:
\begin{eqnarray}
\Gamma\in\{\mathbb{1}, ~\gamma_i, ~\gamma_3, ~\gamma_5, ~i\gamma_i\gamma_5,
~i\gamma_3\gamma_5, ~\sigma_{i3}, ~\epsilon_{ijk}\sigma_{jk}\},
\end{eqnarray}
where $i, j, k\not=3$. The generalized parity and time-reversal operation involve sign flipping of $\delta z$, we define a combination:
\begin{eqnarray}
O_{\Gamma\pm}(\delta z)=
\frac{1}{2}\left\{O_{\Gamma}(\delta z)\pm O_{\Gamma}(-\delta z)\right\},
\label{EQ:nonlocal_bilinear_pm}
\end{eqnarray}
in other words, even/odd function of $\delta z$. The transformation properties for each $\Gamma$ are presented in Table~\ref{TAB:PTCX_nonlocal}. When the chiral symmetry is not imposed, mixings, $(\gamma_{3\pm}\leftrightarrow\mathbb{1}_{\mp})$ and $(\epsilon_{ijk}\sigma_{jk\pm}\leftrightarrow i\gamma_i\gamma_{5\mp})$, are allowed. In a unified way, the mixing can be written as
\begin{eqnarray}
O_{\Gamma\pm}(\delta z)\longleftrightarrow
O_{(1+G_3(\Gamma))\gamma_3\Gamma\mp}(\delta z)\;\;\;\;\;\;
\mbox{(when chiral symmetry is broken)},
\end{eqnarray}
where $G_3(\Gamma)$ is defined to satisfy $\gamma_3\Gamma\gamma_3=G_3(\Gamma)\Gamma$. For the local operator case ($\delta z=0$), this kind of mixing does not occur even when the chiral symmetry is not preserved. It is worthy to note that the separation of two quark fields $\delta z$ acts as ``an extra hand'' to adjust the symmetry transformation property.
\begin{table}[t]
\begin{center}
\begin{tabular}{ccccccccc}
\hline\hline
& $\Gamma=\mathbb{1}_{+/-}$ & $\gamma_{i+/-}$ & $\gamma_{3+/-}$ & $\gamma_{5+/-}$ & $i\gamma_i\gamma_{5+/-}$ & $i\gamma_3\gamma_{5+/-}$ & $\sigma_{i3+/-}$ & $\epsilon_{ijk}\sigma_{jk+/-}$\\
\hline
${\cal P}_3$ & E & O & E & O & E & O & O & E\\       
${\cal P}_{l\not=3}$ & E/O & E/O$_{(l=i)}$ & O/E & O/E & O/E$_{(l=i)}$ & E/O & O/E$_{(l=i)}$ & E/O$_{(l=i)}$\\
&   & O/E$_{(l\not=i)}$ & & & E/O$_{(l\not=i)}$ & & E/O$_{(l\not=i)}$ & O/E$_{(l\not=i)}$\\
${\cal T}_3$ & E/O & E/O & O/E & O/E & O/E & E/O & O/E & E/O\\
${\cal T}_{l\not=3}$ & E & O$_{(l=i)}$ & E & O & E$_{(l=i)}$ & O & O$_{(l=i)}$ & E$_{(l=i)}$\\
& & E$_{(l\not=i)}$ & & & O$_{(l\not=i)}$ & & E$_{(l\not=i)}$ & O$_{(l\not=i)}$\\
${\cal C}$ & E/O & O/E & O/E & E/O & E/O & E/O & O/E & O/E\\
$\chi$ & V & I & I & V & I & I & V & V\\       
\hline\hline
\end{tabular}
\caption{Transformation properties of the nonlocal operator $O_{\Gamma\pm}(\delta z)$: even/odd (E/O) under parity, time reversal, and charge conjugation, and variant/invariant (V/I) under chiral rotation. $i,j,k\not=3$.}
\label{TAB:PTCX_nonlocal}
\end{center}
\end{table}

\section{Symmetry analysis for ${\cal O}(a)$ operators}\label{SEC3}

In this section, we investigate ${\cal O}(a)$ operators for the nonlocal quark bilinear by the symmetry argument used for the analysis on the operator mixing. We prepare a set of ${\cal O}(a)$ higher-dimensional operator for ${\cal O}(pa)$ and ${\cal O}(ma)$:
\begin{eqnarray}
Q_{\Gamma\!\overrightarrow{\rm D}_{\alpha}}(\delta z)&=&
\overline{\psi}(x+\hat{\bm 3}\delta z)
U_3(x+\widehat{\bm 3}\delta z;x)\Gamma\overrightarrow{\slashed{D}}_{\alpha}
\psi(x),
\label{EQ:Opa_non-local_rGD-3}\\
Q_{\overrightarrow{\rm D}_{\alpha}\Gamma}(\delta z)&=&
\overline{\psi}(x+\hat{\bm 3}\delta z)
U_3(x+\widehat{\bm 3}\delta z;x)\overrightarrow{\slashed{D}}_{\alpha}\Gamma
\psi(x),
\label{EQ:Opa_non-local_rDG-3}\\
Q_{\Gamma\!\overleftarrow{\rm D}_{\alpha}}(\delta z)&=&
\overline{\psi}(x+\hat{\bm 3}\delta z)\Gamma
\overleftarrow{\slashed{D}}_{\alpha}U_3(x+\hat{\bm 3}\delta z;x)\psi(x),
\label{EQ:Opa_non-local_lGD-3}\\
Q_{\overleftarrow{\rm D}_{\alpha}\Gamma}(\delta z)&=&
\overline{\psi}(x+\hat{\bm 3}\delta z)\overleftarrow{\slashed{D}}_{\alpha}
\Gamma U_3(x+\hat{\bm 3}\delta z;x)\psi(x),
\label{EQ:Opa_non-local_lDG-3}\\
Q_{\Gamma}^{\rm M}(\delta z)&=&
m\overline{\psi}(x+\hat{\bf 3}\delta z)\Gamma
U_3(x+\hat{\bf 3}\delta z;x)\psi(x),
\label{EQ:Opa_non-local_M}
\end{eqnarray}
where $\alpha\in[3,\perp]$ and we introduce shorthand notations: $\overrightarrow{\slashed{D}}_3=\gamma_3\overrightarrow{D}_3$, $\overrightarrow{\slashed{D}}_{\perp}=\sum_{\mu\not=3}\gamma_{\mu}\overrightarrow{D}_{\mu}$. Due to the same reason as the ${\cal O}(a^0)$ operator, it is convenient to define combinations:
\begin{eqnarray}
Q_{\Gamma\!\overrightarrow{\rm D}_{\alpha}\pm/\overrightarrow{\rm D}_{\alpha}\Gamma\pm/\Gamma\!\overleftarrow{\rm D}_{\alpha}\pm/\overleftarrow{\rm D}_{\alpha}\Gamma\pm}(\delta z)
&=&
\frac{1}{2}\left\{
Q_{\Gamma\!\overrightarrow{\rm D}_{\alpha}/\overrightarrow{\rm D}_{\alpha}\Gamma/\Gamma\!\overleftarrow{\rm D}_{\alpha}\pm/\overleftarrow{\rm D}_{\alpha}\Gamma\pm}(\delta z)\pm
Q_{\Gamma\!\overrightarrow{\rm D}_{\alpha}/\overrightarrow{\rm D}_{\alpha}\Gamma/\Gamma\!\overleftarrow{\rm D}_{\alpha}\pm/\overleftarrow{\rm D}_{\alpha}\Gamma\pm}(-\delta z)
\right\}.
\end{eqnarray}
Also, taking into account the charge conjugation property, we further define combinations,
\begin{eqnarray}
Q_{\Gamma\pm/\overline{\Gamma}\pm}^{{\rm D}_{\alpha}(+)}(\delta z)&=&
Q_{\overleftarrow{\rm D}_{\alpha}\Gamma\pm/\Gamma\!\overleftarrow{\rm D}_{\alpha}\pm}(\delta z)
+Q_{\Gamma\overrightarrow{\rm D}_{\alpha}\pm/\overrightarrow{\rm D}_{\alpha}\!\Gamma\pm}(\delta z),
\label{EQ:Opa_non-local_D+}
\\
Q_{\Gamma\pm/\overline{\Gamma}\pm}^{{\rm D}_{\alpha}(-)}(\delta z)&=&
Q_{\overleftarrow{\rm D}_{\alpha}\Gamma\pm/\Gamma\!\overleftarrow{\rm D}_{\alpha}\pm}(\delta z)
-Q_{\Gamma\overrightarrow{\rm D}_{\alpha}\pm/\overrightarrow{\rm D}_{\alpha}\!\Gamma\pm}(\delta z),
\label{EQ:Opa_non-local_D-}
\end{eqnarray}
which have the properties,
\begin{eqnarray}
Q_{\Gamma\pm/\overline{\Gamma}\pm}^{{\rm D}_{\alpha}(+)}(\delta z)
&\xrightarrow[]{\cal C}&
\mp Q_{(C\Gamma C^{-1})^{\top}\pm/(C\overline{\Gamma}C^{-1})^{\top}\pm}^{{\rm D}_{\alpha}(+)}(\delta z),
\\
Q_{\Gamma\pm/\overline{\Gamma}\pm}^{{\rm D}_{\alpha}(-)}(\delta z)
&\xrightarrow[]{\cal C}&
\pm Q_{(C\Gamma C^{-1})^{\top}\pm/(C\overline{\Gamma}C^{-1})^{\top}\pm}^{{\rm D}_{\alpha}(-)}(\delta z).
\end{eqnarray}
The transformation property of ${\cal O}(pa)$ operators under discrete-symmetries and chiral-symmetry transformations are presented in Table~\ref{TAB:PTCX_nonlocal_Opa}.
\begin{table}[t]
\begin{center}
\begin{tabular}{ccccccccc}
\hline\hline
& $\Gamma=\mathbb{1}_{+/-}$ & $\gamma_{i+/-}$ & $\gamma_{3+/-}$ & $\gamma_{5+/-}$ & $\gamma_i\gamma_{5+/-}$ & $\gamma_3\gamma_{5+/-}$ & $\sigma_{i3+/-}$ & $\epsilon_{ijk}\sigma_{jk+/-}$\\
\hline
${\cal P}_3$ & E & O & E & O & E & O & O & E\\       
${\cal P}_{l\not=3}$ & E/O & E/O$_{(l=i)}$ & O/E & O/E & O/E$_{(l=i)}$ & E/O & O/E$_{(l=i)}$ & E/O$_{(l=i)}$\\
& & O/E$_{(l\not=i)}$ & & & E/O$_{(l\not=i)}$ & & E/O$_{(l\not=i)}$ & O/E$_{(l\not=i)}$\\
${\cal T}_3$ & E/O & E/O & O/E & O/E & O/E & E/O & O/E & E/O\\
${\cal T}_{l\not=3}$ & E & O$_{(i=l)}$ & E & O & E$_{(l=i)}$ & O & O$_{(l=i)}$ & E$_{(l=i)}$\\
& & E$_{(l\not=i)}$ & & & O$_{(l\not=i)}$ & & E$_{(l\not=i)}$ & O$_{(l\not=i)}$\\ 
${\cal C}$($Q_{\Gamma\pm/\overline{\Gamma}\pm}^{{\rm D}_{\alpha}(+)}$) & O/E & E/O & E/O & O/E & O/E & O/E & E/O & E/O\\
${\cal C}$($Q_{\Gamma\pm/\overline{\Gamma}\pm}^{{\rm D}_{\alpha}(-)}$) & E/O & O/E & O/E & E/O & E/O & E/O & O/E & O/E\\
$\chi$ & I & V & V & I & V & V & I & I\\       
\hline\hline
\end{tabular}
\caption{Transformation properties of the nonlocal operators at ${\cal O}(pa), $ $Q_{\Gamma\pm/\overline{\Gamma}}^{{\rm D}_{\alpha}(\pm)}(\delta z)$: even/odd (E/O) under parity, time reversal and charge conjugation, and variant/invariant (V/I) under chiral rotation. $i,j,k\not=3$.}
\label{TAB:PTCX_nonlocal_Opa}
\end{center}
\end{table}
By comparing this with Table~\ref{TAB:PTCX_nonlocal}, we can determine which higher-dimensional operators are allowed as the ${\cal O}(a)$ of $O_{\Gamma}(\delta z)$. When we do not impose the chiral symmetry, we observe that each ${\cal O}(a^0)$ operator, $O_{\Gamma}(\delta z)$, can have ${\cal O}(pa)$ operators with two $\Gamma$s, $Q_{\Gamma'=\Gamma}^{{\rm D}_{\alpha}}(\delta z)$, and $Q_{\Gamma'=\gamma_3\Gamma}^{{\rm D}_{\alpha}}(\delta z)$.

We summarize the operator mixing on the nonlocal quark bilinear and possible ${\cal O}(a)$ operators. Without the chiral symmetry, the ${\cal O}(a^0)$ operator $O_{\Gamma\pm}(\delta z)$ in Eq.~(\ref{EQ:nonlocal_bilinear_pm}) can mix with
\begin{eqnarray}
O_{\chi^B\Gamma\pm}(\delta z)=(1+G_3(\Gamma))O_{\gamma_3\Gamma\mp}(\delta z),
\end{eqnarray}
where a subscript $\chi^B$ indicates the effect of chiral-symmetry breaking. Having definitions in Eqs.~(\ref{EQ:Opa_non-local_M}), (\ref{EQ:Opa_non-local_D+}), and (\ref{EQ:Opa_non-local_D-}), possible ${\cal O}(a)$ operators are
\begin{eqnarray}
O_{\Gamma\pm}^{{\cal O}(p_{\alpha}a)}(\delta z)
&=&(1+G_3(\Gamma))Q_{\gamma_3\Gamma\mp}^{{\rm D}_{\alpha}(-)}(\delta z)
+(1-G_3(\Gamma))O_{\gamma_3\Gamma\mp}^{{\rm D}_{\alpha}(+)}(\delta z),
\label{EQ:nonlocal-Opa1}
\\
O_{\overline{\Gamma}\pm}^{{\cal O}(p_{\alpha}a)}(\delta z) 
&=&(1+G_3(\Gamma))
Q_{\overline{\gamma_3\Gamma}\mp}^{{\rm D}_{\alpha}(-)}(\delta z)
+(1-G_3(\Gamma))
O_{\overline{\gamma_3\Gamma}\mp}^{{\rm D}_{\alpha}(+)}(\delta z),
\label{EQ:nonlocal-Opa2}
\\
O_{\Gamma\pm}^{{\cal O}(ma)}(\delta z) 
&=&(1+G_3(\Gamma))Q_{\gamma_3\Gamma\mp}^{\rm M}(\delta z),
\label{EQ:nonlocal-Oma}
\\
O_{\chi^B\Gamma\pm}^{{\cal O}(p_{\alpha}a)}(\delta z) 
&=&Q_{\Gamma\pm}^{{\rm D}_{\alpha}(-)}(\delta z),
\label{EQ:nonlocal-Opa1_cb}
\\
O_{\chi^B\overline{\Gamma}\pm}^{{\cal O}(p_{\alpha}a)}(\delta z) 
&=&Q_{\overline{\Gamma}\pm}^{{\rm D}_{\alpha}(-)}(\delta z),
\label{EQ:nonlocal-Opa2_cb}
\\
O_{\chi^B\Gamma\pm}^{{\cal O}(ma)}(\delta z) 
&=&Q_{\Gamma\pm}^{\rm M}(\delta z).
\label{EQ:nonlocal-Oma_cb}
\end{eqnarray}
Among them, $O_{\overline{\Gamma}\pm}^{{\cal O}(p_3a)}$ and $O_{\chi^B\overline{\Gamma}\pm}^{{\cal O}(p_3a)}$ are redundant. Notably, there are ${\cal O}(a)$ contributions even when chiral fermions are employed, which is quite different from the local-operator case.

\section{Summary}\label{SEC4}

In these proceedings, the one-loop perturbative continuum-lattice matching for the nonlocal quark bilinear, which appears in the quasi-PDF approach, was demonstrated with Wilson quark formalism. We have also investigated the operator mixing for a class of nonlocal operator (\ref{EQ:nonlocal_bilinear}) on the lattice using action symmetries: parity, time reversal, charge conjugation, and chiral symmetry. In the discussion, we found that the symmetric and anti-symmetric combination with respect to the separation of quark fields in Eq.~(\ref{EQ:nonlocal_bilinear_pm}) makes the symmetry transformation property more visible. Switching between the symmetric and anti-symmetric combination acts as an ``extra hand'' to adjust the transformation property, and enables the mixing which we cannot see in the local-operator case. The symmetry argument shows that unlike local bilinears, the chiral symmetry is so crucial to prevent the mixing. We also have shown the possible ${\cal O}(a)$ operators for the nonlocal bilinear by extending the symmetry discussion for the ${\cal O}(a^0)$ operator. The important finding is that a part of the ${\cal O}(a)$ operators cannot be prohibited from emerging by the chiral symmetry.

In the quasi-PDF method, we have to use large hadron momenta to control higher-twist contamination. The inclusion of high momentum in the numerical simulations is challenging because signal-to-noise ratio gets worse as the momentum becomes larger. This difficulty could be overcome by using the momentum-smearing technique presented in Ref.~\cite{Bali:2016lva}. However, the large momentum would cause significant lattice discretization errors. To reduce the discretization errors, implementing ${\cal O}(a)$-improvement program would be demanded. Determination of the ${\cal O}(a)$-improvement coefficients for the nonlocal quark bilinear using the one-loop lattice perturbation, and possibly nonperturbative approach, is to be addressed.

\section*{Acknowledgments}

T.I. is supported by Science and Technology Commission of Shanghai Municipality (Grants No. 16DZ2260200) and in part by the Department of Energy, Laboratory Directed Research and Development (LDRD) funding of BNL, under contract DE-EC0012704.



\end{document}